\documentclass[
 wlscirep,
 amsmath,amssymb,
 floatfix,
 notitlepage,
]{revtex4-1}

\usepackage{graphicx}
\usepackage{dcolumn}
\usepackage{bm}
\usepackage{gensymb}
\usepackage[english]{babel}

\begin{document}

\title{Describing transport in defected nanoparticle solids using a new, hierarchical, simulation tool, TRIDENS}

\author{Chase Hansen$^{1,}$\footnote[1]{Correspondence to cmhansen@ucdavis.edu}, Davis Unruh$^1$, Miguel Alba$^1$, Caroline Qian$^2$, Alex Abelson$^2$, Matt Law$^2$, and Gergely T. Zimanyi$^1$}
\affiliation{$^1$Physics Department, University of California, Davis}
\affiliation{$^2$Chemistry Department, University of California, Irvine}

\date{\today}

\begin{abstract}
The efficiency of nanoparticle (NP) solar cells has grown impressively in recent years, exceeding 16\%. However, the carrier mobility in NP solar cells, and in other optoelectronic applications remains low, thus critically limiting their performance. Therefore, carrier transport in NP solids needs to be better understood to further improve the overall efficiency of NP solar cell technology. However, it is technically challenging to simulate experimental scale samples, as physical processes from atomic to mesoscopic scales all crucially impact transport. To rise to this challenge, here we report the development of TRIDENS: the \textbf{Tr}ansport \textbf{i}n \textbf{De}fected \textbf{N}anoparticle \textbf{S}olids Simulator, that adds three more hierarchical layers to our previously developed HINTS code for nanoparticle solar cells. In TRIDENS, we first introduced planar defects, such as twin planes and grain boundaries into individual NP superlattices (SLs) that comprised the order of $10^3$ NPs. Then we used HINTS to simulate the transport across tens of thousands of defected NP SLs, and constructed the distribution of the NP SL mobilities with planar defects. Second, the defected NP SLs were assembled into a resistor network with more than $10^4$ NP SLs, thus representing about $10^7$ individual NPs. Finally, the TRIDENS results were analyzed by finite size scaling to explore whether the percolation transition, separating the phase where the low mobility defected NP SLs percolate, from the phase where the high mobility undefected NP SLs percolate drives a low-mobility-to-high-mobility transport crossover that can be extrapolated to genuinely macroscopic length scales. For the theoretical description, we adapted the Efros-Shklovskii bimodal mobility distribution percolation model. We demonstrated that the ES bimodal theory's two-variable scaling function is an effective tool to quantitatively characterize this low-mobility-to-high-mobility transport crossover. 

\begin{description}

\item[PACS numbers] 73.63.Kv

\item[Keywords] nanoparticle, quantum dot, commensuration, FET, transport
\end{description}
\end{abstract}

\maketitle

Colloidal semiconductor nanoparticles (NPs) are singularly promising nanoscale building blocks for fabricating mesoscale materials that exhibit emergent collective properties. There is a growing interest to use NPs for numerous optoelectronic applications \cite{talapin_prospects_2010,doi:10.1021/nn506223h}, including third generation solar cells\cite{Nozik02,doi:10.1021/jp806791s} light emitting diodes\cite{shirasaki2013emergence}, and field effect transistors (FET)\cite{Talapin07102005,hetsch_quantum_2013}.

Electron wavefunctions are localized on the individual NPs. This ``quantum confinement" makes the electronic parameters tunable with the NP size, and thus makes the NP solids a very versatile platform for applications\cite{voros2015colloidal}. However, the very same quantum confinement also suppresses the transport between NPs, and thus drives NP solids insulating. As a result, without the application of specific transport-boosting fabrication steps, the electron mobility in NP solids is often in the range of $10^{-2}-10^{-1}$ cm$^2$/Vs\cite{choi_bandlike_2012, MurrayKagan}. These mobilities are typically measured in FET arrangements. This is orders of mangitude below the mobilities that would be acceptable for electronic applications. Therefore, increasing the mobility and transport in NP solids is one of the central challenges on the way to realize the promise of NP solids.

Various experimental groups managed to boost the mobility by enhancing the inter-NP transition rate with a variety of methods, including: ligand engineering \cite{wang2016colloidal,jang_temperature-dependent_2014,lee_band-like_2011}, band-alignment engineering\cite{chuang2014improved,kroupa2017tuning}, chemical-doping\cite{chen2016metal,choi_bandlike_2012}, photo-doping \cite{talgorn2011unity}, metal-NP substitution\cite{cargnello2015substitutional}, epitaxial attachment of NPs \cite{doi:10.1021/acs.nanolett.6b02382,whitham2016charge}, and atomic layer deposition methods\cite{liu2013pbse}. Encouragingly, these efforts recently translated into notable progress, as NP solids were reported to exhibit band-like, temperature-insensitive mobilities, with values exceeding 10 ${\rm cm^2/Vs}$ at room temperatures \cite{choi_bandlike_2012, MurrayKagan}. It is important to note that some experiments reported data that can be interpreted as evidence for band-like transport. One of these is the relative temperature independence of the observed mobilities, in contrast to hopping insulators where an activated temperature dependence is expected. However, the absolute values of the mobilities remain relatively low compared to most metals, and this makes conservative commentators stop short of identifying this transport as metallic\cite{choi_bandlike_2012, MurrayKagan}.

On the theoretical front, there have been efforts from several groups to understand electronic transport in NP films and solids. Density functional theory (DFT)-based ab initio calculations of the energy levels of a single NP alone are already limited to only hundreds of atoms for higher-reliability methods, and a few thousands for more approximate methods by prohibitive CPU times. These translate to diameters less than 2-3 nm, whereas experimental NP diameters often exceed 5-6 nm. Next, the accurate computation of the NP-NP transition rates would require the simulation of two NPs. And even if this calculation is completed, it does not address that the NP-NP transport is not metallic but insulating; the disorder of the parameters from NP to NP; and finally the defects of the NP solids. In total, ab initio descriptions alone are very far from being capable of describing transport in NP solids. Cleary, there is a pressing need for developing mesoscopic transport simulations that somehow integrate ab initio calculations.

Shklovskii et al. have developed transport calculations for a NP array in a FET geometry, where they focused on the effects of the Coulomb interaction \cite{reich2014theory}. The interplay of transport and Coulomb interactions was studied in Refs. \citenum{PhysRevB.89.235303} and \citenum{chandler_electron_2007}, albeit on very small samples. Over the last few years, our group developed the Hierarchical Nanoparticle Transport Simulator (HINTS) platform that starts with an ab initio calculation of the energetics of individual nanoparticles, then forms a NP solid of several hundred NPs, and finally simulates transport across this NP solid by a Kinetic Monte Carlo method\cite{carbone_monte_2013,qu2017metal}. HINTS can simulate 500-2,000 nanoparticles. A reassuring validation of HINTS emerged from simulating the dependence of the mobility of PbSe NP layers as a function of the NP diameter. The results in \cite{carbone_monte_2013,qu2017metal} closely tracked the experimental results of Liu et al., who studied the electron mobility of PbSe layers in a FET geometry \cite{liu_dependence_2010}. More recently, we studied commensuration effects in bilayer NP solids\cite{PhysRevB.101.045420}.

\begin{figure*}[t]
\centering
    \includegraphics[width=\textwidth]{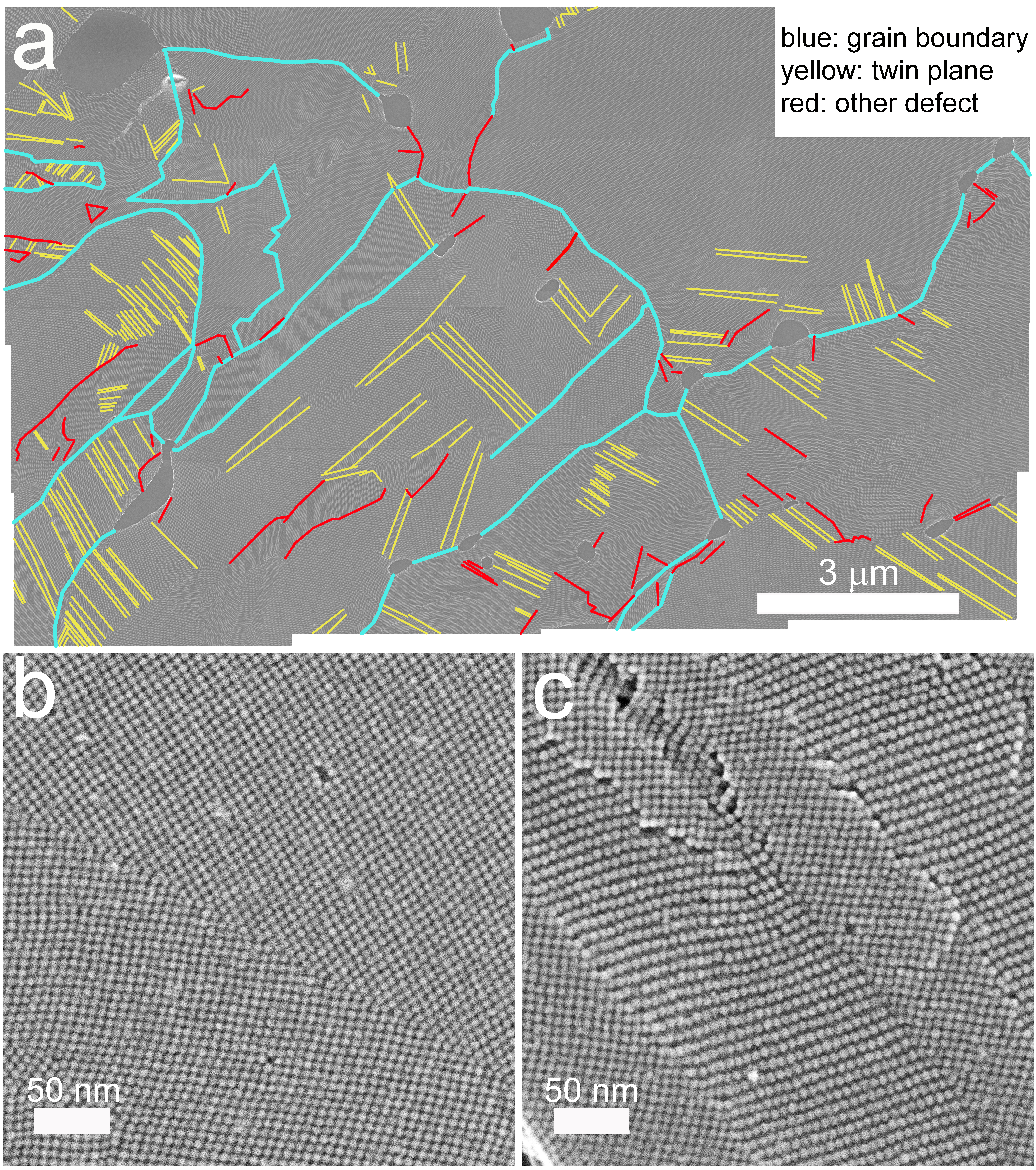}
    \caption{Common planar defects in PbSe nanoparticle superlattices. (a) Grain map of a typical region of an PbSe NP superlattice film showing the location of several types of planar defects. Blue, yellow, and red lines denote wide-angle grain boundaries, twin planes, and more complex, unclassified planar defects and defect clusters, respectively. The image is a montage of fifteen low-magnification, high-resolution SEM images. Voids, step edges, vacancies, and other types of non-planar defects are also visible in the montage. (b) Higher-magnification secondary electron image of a bicrystalline region of an epi-SL film with two (100)SL-oriented SL grains meeting at a twin plane. (c) Image of another region of the same epi-SL film showing multiple grain boundaries between (100)SL- and (01${\bar{} 1}$)SL-oriented grains, as well as several other planar defects.}
    \label{sem}
\end{figure*}

However, these theoretical efforts only considered NP solids with homogeneous disorder: the NPs were arranged either in a close-packed glassy/jammed structure, or on an ordered superlattice (SL) with disorder only in the NP size. In contrast, representative scanning electron microscope (SEM) images, like in Fig. \ref{sem}, taken of NP solids with millions of NPs, conspicuously reveal that typical NP solids are also characterized by disorder on much larger length scales. These defects, often on the $\mu$m length scale, have sizes well beyond the capabilities of any published technique, including HINTS. Therefore, there is a need for transport simulation methods that are capable of capturing meso- and macro-scale defects and their effect on transport.

We performed one step in this direction previously by extending our HINTS method to include percolative effects into homogeneously disordered NP solids\cite{Qu2017}. This simulation captured physics on the longer length scales of percolative clusters. Our main message was that a metal-insulator transition (MIT) occurs when a percolating path of metallic-connected NPs develops across the entire sample. We described this MIT as a Quantum Percolation Transition. However, this work still did not incorporate planar defects.

\begin{center}
\underline{Simulation Methods}
\end{center}

To answer the above needs, in this paper we report our work that boosted the capability of our HINTS platform by introducing additional hierarchical layers to capture the effect of planar defects on the transport in NP solids. First, we used HINTS to individually model a NP superlattice (SL) with one planar defect that was either a generic grain boundary or a twin plane. Second, we simulated transport across a large number of such single-defect SLs, and determined the distribution of the mobilities of the single-grain-boundary NP SLs and that of the single twin-plane NP SLs. We also determined the distribution of the mobilities of undefected NP SLs with only homogeneous NP disorder. Third, to reach a simulation scale approaching the scale of the NP solids in the experiments, we built a resistor network where the individual resistor values were taken from the three mobility distributions with predetermined probabilities. Motivated by our previous work\cite{Qu2017}, we determined the resistance of the entire resistor network by changing the fraction of undefected NP SLs within the network. Finally, we analyzed our results by a finite size scaling method.

We call this boosted HINTS platform the TRIDENS: the ``\textbf{TR}ansport \textbf{I}n \textbf{DE}fected \textbf{N}anoparticle \textbf{S}olids” Simulator. With TRIDENS, we are capable of capturing the physics from atomistic length scales up to the scale of NP solids in the experiments by integrating the simulations on several hierarchical layers. The complete hierarchical structure of TRIDENS is presented below.

(1) The energy levels of individual PbSe NPs are determined by adapting a $k \cdot p$ calculation within the NP diameter range of 5-7 nm. The valence band and conduction band values have been validated via comparison to optical experiments \cite{kang_electronic_1997}. Here we focus on PbSe NPs because they are of considerable interest for solar applications due to their large Bohr exciton radius and small direct bulk bandgap \cite{Hostetler}, and exhibit the possibly game-changing multiple-exciton generation (MEG)\cite{Klimov}. For these reasons, PbSe NPs are often thought to have strong promise for solar applications. 

(2) The electron-electron interaction is included on the level of an on-site, or self-charging energy expression, $E_C$, defined as:
\begin{equation}
    E_C = n(\Sigma^0 + \frac{(n-1)}{2}\Sigma)
\end{equation}
where $n$ is the number of electrons on the NP. $\Sigma^0$ is the self-charging energy of loading the first electron onto a neutral NP. $\Sigma$ is the extra energy it takes to load each additional electron onto the NP due to repulsive Coulomb interaction with the (n-1) electrons already on the NP, as well as the interaction with the induced image charge. 

This self-charging energy can be calculated by a variety of methods, including the semi-empirical pseudopotential configuration interaction method of Zunger and coworkers \cite{Zunger-charging} and the single NP empirical-perturbative hybrid calculations of Delerue \cite{delerue}. In this paper we report results with the latter approach. In this approach
\begin{equation}
    \Sigma^0 = \frac{q^2}{8\pi \epsilon_0 R}(\frac{1}{\epsilon_{\mathrm{solid}}} - \frac{1}{\epsilon_{\mathrm{NP}}}) + 0.47\frac{q^2}{4\pi \epsilon_0 \epsilon_{\mathrm{NP}} R}(\frac{\epsilon_{\mathrm{NP}} - \epsilon_{\mathrm{solid}}}{\epsilon_{\mathrm{NP}} + \epsilon_{\mathrm{solid}}}),
\end{equation}
and
\begin{equation}
    \Sigma = \frac{q^2}{4\pi \epsilon_0 R}(\frac{1}{\epsilon_{\mathrm{solid}}} + 0.79\frac{1}{\epsilon_{\mathrm{NP}}}).
\end{equation}
For the dielectric constant inside the NP, we assume that it equals the bulk high frequency dielectric constant of PbSe, taken to be 22.0. To model the dielectric constant of the medium surrounding the NP, we account for both the organic ligand shell of the NP as well as the presence of neighboring NPs. We assume that the ligands themselves have a dielectric constant of 2.0. The dielectric constant of the entire solid is then calculated using the Maxwell-Garnett (MG) effective medium approximation:
\begin{equation}
    \epsilon_{\mathrm{solid}} = \epsilon_{\mathrm{ligand}}\frac{\epsilon_{\mathrm{NP}}(1 + \kappa f) - \epsilon_{\mathrm{ligand}}(\kappa f - \kappa)}{\epsilon_{\mathrm{ligand}}(\kappa + f) + \epsilon_{\mathrm{NP}}(1-f)}
\end{equation}
where $\kappa$ is 2 for spherical NPs, and $f$ is the filling factor.

We note that the long range part of the Coulomb interaction can be easily included into the calculation. The long range interactions change the nature of transport from activated hopping to Efros-Shklovski type variable range hopping. However, many experiments show that while this Efros-Shklovskii hopping dominates at low temperatures, as the temperature is raised past 150-200K, the transport becomes dominated by nearest neighbor hopping.\cite{doi:10.1021/jz300048y} Since our work focuses on temperature ranges around ambient room temperature, representing the Coulomb interactions with the on-site term only is appropriate.

(3) We modelled the electron transitions between neighboring NPs via a Miller-Abrahams phonon-assisted hopping mechanism: 
\begin{equation}
    \Gamma_{i\rightarrow j} = \begin{cases} \nu g_{ij} \beta_{ij} \exp{\frac{-\Delta E_{ij}}{k_b T}} & \textrm{if } \Delta E_{ij} \textrm{ }> \textrm{ }0
    \\
    \nu g_{ij} \beta_{ij} & \textrm{if } \Delta E_{ij} <= 0 \end{cases}
\end{equation}
where $\nu$ is an attempt frequency, chosen to be $10^{12} \, \mathrm{s}^{-1}$, $g_{ij}$ is the product of the initial density of states on NP$_i$ and the final density of states on NP$_j$, and $\beta_{ij}$ is the tunneling amplitude. $\beta_{ij}$ is calculated using the WKB approximation as:
\begin{equation}
    \beta_{ij} = \exp{(-2\Delta x \sqrt{\frac{2m^* (E_{\mathrm{vac}}-E_{ij})}{\hbar^2}})}
\end{equation} 
Here $\Delta x$ is the NP-NP surface-surface separation distance. $m^*$ is the effective mass of the electrons in the tunneling medium, approximated as .05$m_e$, the effective mass of electrons in bulk PbSe. It is noted that $m^*$ was estimated to be 0.3$m_e$ in NPs\cite{doi:10.1021/nl101284k}. $E_{\mathrm{vac}}$ is the vacuum energy level, set to be zero as all other energy levels are defined relative to the vacuum. $E_{ij}$ is the tunneling energy, taken to be the average of the initial and final states of the hopping transition: $E_{ij} = (E_i + E_j)/2$, where $E_i$ is the energy level of $NP_i$.

\begin{figure*}[t]
\includegraphics[width=\textwidth]{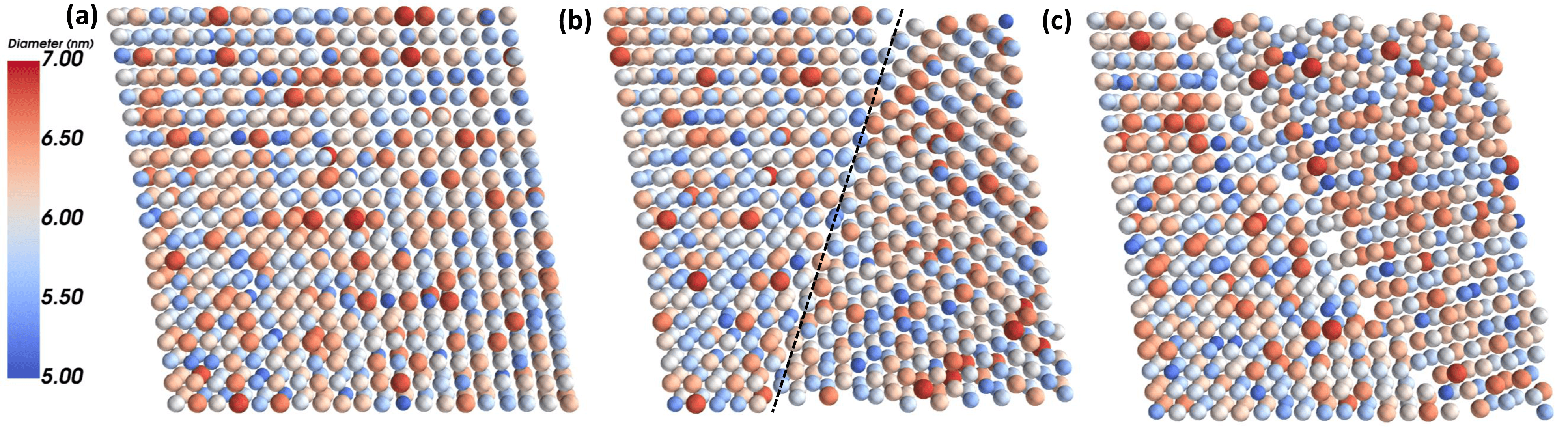}
\caption{Top down views of the three types of the simulated NP SLs: (a) An undefected NP SL, characterized by the NPs having only size and location disorder; (b) A NP SL, containing a twin plane, as denoted by the dashed line, also with NP size and location disorder; and (c) An NP SL, containing a grain boundary, also with NP size and location disorder. NP color corresponds to NP diameter, as indicated in the colorbar on the left.}
\label{visualization}
\end{figure*}

(4) To reach the length scale of hundreds of nanometers, we generated triclinic NP superlattices, of PbSe NPs with a 20x20x2 geometry, inspired by the 2D channel geometries of FETs used in transport experiments \cite{ISI:000511170100013}. The triclinic unit cell was described with lattice constants $a_1$ = $a_2$ = $a_3$ = 6.9 nm and angles $\alpha$ = $\beta$ = $\gamma$ = 99$\degree$. The average NP diameter was 6.0 nm. Size and location disorder were introduced by assigning the NPs a diameter and lattice displacement vector according to Gaussian distributions of widths $\sigma (diameter)= 0.4$ nm and $\sigma (location)= 0.3$ nm respectively. See Fig. \ref{visualization}a for an example of one of these SLs. In our undefected SLs, these parameters yield a $\langle\beta_{ij}\rangle \, \simeq \, 0.015 \pm .02$.

Layers (1)-(4) are the main constituents of our HINTS platform. HINTS is suitable for capturing the effects of homogeneous disorder, i.e. disorder associated with the size and location of the NPs that varies from site to site of the NP superlattice, but does not involve planar defects. Next, we describe the additional layers of the TRIDENS that enable us to access length scales well beyond the reach of HINTS.

(5) As a first step, we introduced a single planar defect into each generated NP superlattice (SL). We carefully analyzed SEM images of NP solids with millions of NPs of the type of Fig.\ref{sem}, and determined the predominant types of defects and their statistics, such as the lengths and densities of the planar defects. Based on the SEM image analysis, the most relevant and oft-occurring planar defects were twin planes and grain boundaries.

(a) Twin planes: the NP superlattice is mirrored across a boundary plane, creating two crystallites, or grains, which have reflected in-plane unit cells. Twinned grains can also be related by a $180\degree$ rotation normal to the twin plane. The twin plane is always parallel to a possible crystal face (but not any planes of complete symmetry, e.g. it is distinct from all space group symmetries), and thus requires the two grains to share some NP lattice sites along the boundary plane. See Fig. \ref{sem}b and Fig. \ref{visualization}b. The high symmetry nature of twin planes makes them a low disorder defect, compared to the more highly disordered grain boundaries discussed below. In our generated samples the twin planes were created in a (100) in-plane oriented SL, and the orientation of the twin plane itself was randomly selected on a sample-by-sample basis from all possible crystal planes which would span the entire NP simulation SL in the x-direction (in order to bisect the SL). As an example, one such boundary orientation is that of a $(01\bar{2})/(02\bar{1})$ twin boundary, where $(01\bar{2})$ is the orientiation of the boundary plane in grain 1, and $(02\bar{1})$ is the orientiation of the boundary plane in the mirrored grain 2.

(b) Grain boundaries: the NP superlattice is again fractured by a boundary plane. However, unlike with twin planes, the superlattice is not mirrored across the boundary plane. There are two main types of grain boundaries. 1) Tilt grain boundaries, where the in-plane SL orientation is the same in the two grains, but they are spatially rotated in-plane relative to each other. The angle of rotation can be divided into ``low angle" and ``high angle" regimes, where the higher the angle of rotation, the more disordered the grain boundary (with large areas of poor fit). 2) Twist grain boundaries, where the in-plane superlattice orientations of the two grains are different (rotation occurs along an axis perpendicular to the boundary plane). Such grain boundaries will result in two crystallites/grains with different in-plane superlattice orientations (e.g. a boundary between a $(100)_{\mathrm{SL}}$ in-plane orientation and a $(101)_{\mathrm{SL}}$ in-plane orientation). 

In our generated samples, we simulated grain boundaries with a combination of tilt and twist mismatching. Specifically, the boundary plane separated grains of $(100)_{\mathrm{SL}}$ in-plane orientation and $(01\bar{1})_{\mathrm{SL}}$ in-plane orientation respectively, with the relative in-plane spatial orientation of the two grains depending on the angle of the boundary plane (chosen at random on a sample-by-sample basis). The boundary plane was always limited to angles which would span the entire NP simulation SL in the x-direction (in order to bisect the SL). This results in grain boundaries which are much more extensively disordered than twin planes, particularly when the boundary plane results in a high-tilt grain boundary. See Fig. \ref{sem}c and Fig. \ref{visualization}c. Hereafter, we will refer to our specific combination of boundary mismatching as simply a ``grain boundary".

In total, we generated 30,000 NP superlattices, containing either one or two grains, where we varied the disorder of the NP diameters (see color code in Fig. 2), the on-site NP location disorder, and the orientation of the planar defects as viewed out-of-plane. Of these 30,000 NP SLs, 10,000 NP SLs had no planar defects, the next 10,000 NP SLs contained one twin plane, and the last 10,000 NP SLs contained a grain boundary. 

Fig.\ref{sem} shows other types of defects as well. We determined that point vacancies have minimal effect on the mobilities in the insulating phase. One can also see tears/rips/voids/cracks in the SEM image. NP superlattice fabrication technologies will be ready for technical application when they can minimize or eliminate such disruptive tears. For these reasons, we did not model either of these defects.

(6) Next, we determined the electron mobility across each of the 3x10,000 defected NP SLs. To do this, each NP SL was populated with electrons, randomly placing them on NPs, using the Mersenne Twister, until a predetermined electron density was reached. The chance of an electron being placed on any particular NP was uniform, independent of electron occupation and NP parameters. Data was only taken well after the system achieved equilibrium. A small voltage was applied across the sample to induce transport in the linear I-V regime, with periodic boundary conditions. Finally, the electron transport was simulated by evolving time via a kinetic Monte Carlo (KMC) algorithm. The so-determined mobilities of the 3x10,000 NP SLs were used to create the mobility distributions for the homogeneously disordered NP SLs, the twin-plane-supporting NP SLs, and the grain-boundary-supporting NP SLs. The first class of NP SLs will also be referred to as ``undefected NP SLs'', the latter two classes as ``defected NP SLs".

(7) To simulate NP solids on mesoscopic length scales of the order of 10 $\mu$m or longer, we generated a classical resistor network, with resistors chosen at random from the distributions determined in step 6. Which distribution the resistors were chosen from was also randomly determined, according to a parameter that describes the fraction of defected resistors. Random numbers were generated using standard numpy libraries. Each resistor represents an NP SL with an $L=20$ length planar defect. One notes that in Fig.\ref{sem} many of the planar defects are considerably longer than $L=20$. Representing planar defects with longer lengths is possible in TRIDENS by placing defected NP SLs correlated along the lines of the network. Such longer range defect-correlations were not pursued in the present work, but will be included in future work.

With these preparations, the mobility of the overall NP solid was determined by treating this NP SL network as a resistor network. We used the Laplacian method of F. Y. Wu et al. to calculate the overall resistance across the entire network \cite{Wu_2013}. The electrodes were modeled as equipotential metallic strips spanning the entire length of the sample edge, thus making them equivalent to a single node on a resistor network. These electrodes were coupled to the sample by contact resistors that were chosen according to the same rules as the bulk resistors.

(8) Having determined the overall mobility of the network of defected NP SLs, we adapted finite size scaling methods to analyze whether this resistor network model built from defected NP SLs had a phase transition, or a crossover, and if so, what are the properties of this transition. To this end, we repeated step (7) for resistor networks of various sizes, including 32x32, 64x64, and 128x128. As detailed below, our finite size scaling found a percolation transition that separates a low mobility insulator from a high mobility insulator. We used finite size scaling to determine the critical properties of this transition, including the critical point, the critical exponents and the universal scaling function.

\begin{center}
\underline{Experimental Methods}
\end{center}

\underline{Materials:}
Lead oxide (PbO, 99.999\%), oleic acid (OA, technical  grade, 90\%), diphenylphosphine (DPP, 98\%), 1-octadecene (ODE, 90\%), ethylene glycol (EG, 99.8\%, anhydrous), acetonitrile (99.99\%, anhydrous), hexanes ($\geq$99\%, anhydrous), toluene (99.8\%, anhydrous), and (3-mercaptopropyl)trimethoxysilane (3-MPTMS, 95\%) were purchased from Sigma Aldrich and used as received. Trioctylphosphine (TOP, technical grade, $>$90\%) and selenium (99.99\%) were acquired from Fluka and mixed for 24 hours to form a 1 M TOP-Se stock solution. Ethylenediamine (EDA,  $>$98.0\%, anhydrous) was purchased from TCI and mixed with acetonitrile in a 1:1 volume ratio to make a 7.5 M EDA stock solution, this is a slight modification to a published procedure\cite{ISI:000511170100013}.

\underline{Quantum Dot Synthesis:}
In this experimental section we adopt the alternative terminology of ``quantum dots" to refer to the nanoparticles, to accommodate alternative terminologies preferred by different communities. PbSe QDs were synthesized and purified air-free using a slight modification of a published procedure \cite{ISI:000511170100013}. Briefly, PbO (1.50 g), OA (5.00 g), and ODE (10.00 g) were mixed and degassed in a three-neck round-bottom flask at room temperature.  The mixture was heated to 120 C under vacuum to form dissolved Pb(OA)2 and dry the solution. After 1 hour at 120 C, the Pb(OA)2 solution was heated to 180 C under argon flow and 9.5 mL of a 1 M solution of TOP-Se containing 200 $\mu$L of DPP was rapidly injected into this hot solution. An immediate darkening of the solution was observed, and the QDs were grown for 105 seconds at $\sim$160 C. The reaction was quenched with a liquid nitrogen bath and injection of 10 mL of anhydrous hexanes. QD purification and SL fabrication were performed in glove boxes with $<$0.5 ppm O2 content. The QDs were purified by two rounds of precipitation/redispersion using acetonitrile/toluene and stored as a powder in the glove box.

\underline{Substrate preparation:}
Following and slightly modifying the procedure seen in \cite{ISI:000511170100013} a single-side polished Si substrate was cleaned using 10 minutes of sonication in acetone, Millipore water, and then isopropanol, followed by drying in a stream of flowing air. The cleaned substrate was immersed in a 100 mM solution of 3-MTPMS in toluene for 1 hour to functionalize its native SiOx surface for improved QD film adhesion, then rinsed with neat toluene and dried in flowing air.

\underline{Superlattice fabrication, electron microscopy imaging:}
Quantum dot superlattice films were fabricated and imaged using (modified) published procedures \cite{ISI:000511170100013}. An oleate-capped superlattice was prepared in a glovebox ($<$2 ppm O2) by drop casting 60 $\mu$L of 20 g/L dispersion of PbSe QDs in hexanes onto 7 mL of ethylene glycol (EG) in a Teflon well (3.5 x 5 x 1 cm). After depositing the QD solution, the well was immediately covered with a glass lid. The hexane evaporated over 30 minutes, resulting in a smooth, dry QD film floating on the EG surface. The glass lid was then removed and 0.1 mL of a 7.5 M solution of EDA in acetonitrile was slowly injected (5-10 sec) into the EG under the QD film using a 1 mL syringe.  After 30 seconds of exposure to EDA, the resulting epi-SL film was stamp transferred to the Si substrate using a vacuum wand, rinsed vigorously with acetonitrile, and dried under flowing N2.

Scanning electron microscopy (SEM) imaging was performed on an FEI Magellan 400 XHR SEM operating at 10 kV. Grain maps were produced by stitching together fifteen 6,144 x 4,415 pixel images acquired at 50,000x magnification, providing the ability to resolve individual QDs in the sample. Image stitching was performed in Adobe Photoshop. Grain boundaries, twin planes, and other planar defects were then located by eye and drawn in manually.

Superlattice samples for TEM analysis were prepared by stamping QD films from the EG surface onto holey carbon TEM grids without a carbon film coating. The use of TEM grids free of a carbon film was critical for high-quality secondary electron imaging (SEI) in the TEM. SE imaging was performed on a JEOL JEM-2800 TEM operating at 200 kV using a probe size of 1.0 nm.

\begin{center}
\underline{Results and Discussion}  
\end{center}

\underline{TRIDENS simulations:} We laid the foundation of our simulation by carrying out steps (1)-(4), as done in a standard HINTS study. To carry out step (5), we generated 3x10,000 defected NP SLs by starting with homogeneously disordered but undefected 20x20x2 NP SLs whose shape broadly corresponded to FET geometries, and then inserted a twin plane planar defect into 10,000 NP SLs, and a grain boundary planar defect into another 10,000 NP SLs. The latter sometimes involved removing a few NPs to keep the shape of the NP SLs largely unchanged. 

\begin{figure}[h]
\includegraphics[width=0.5\columnwidth]{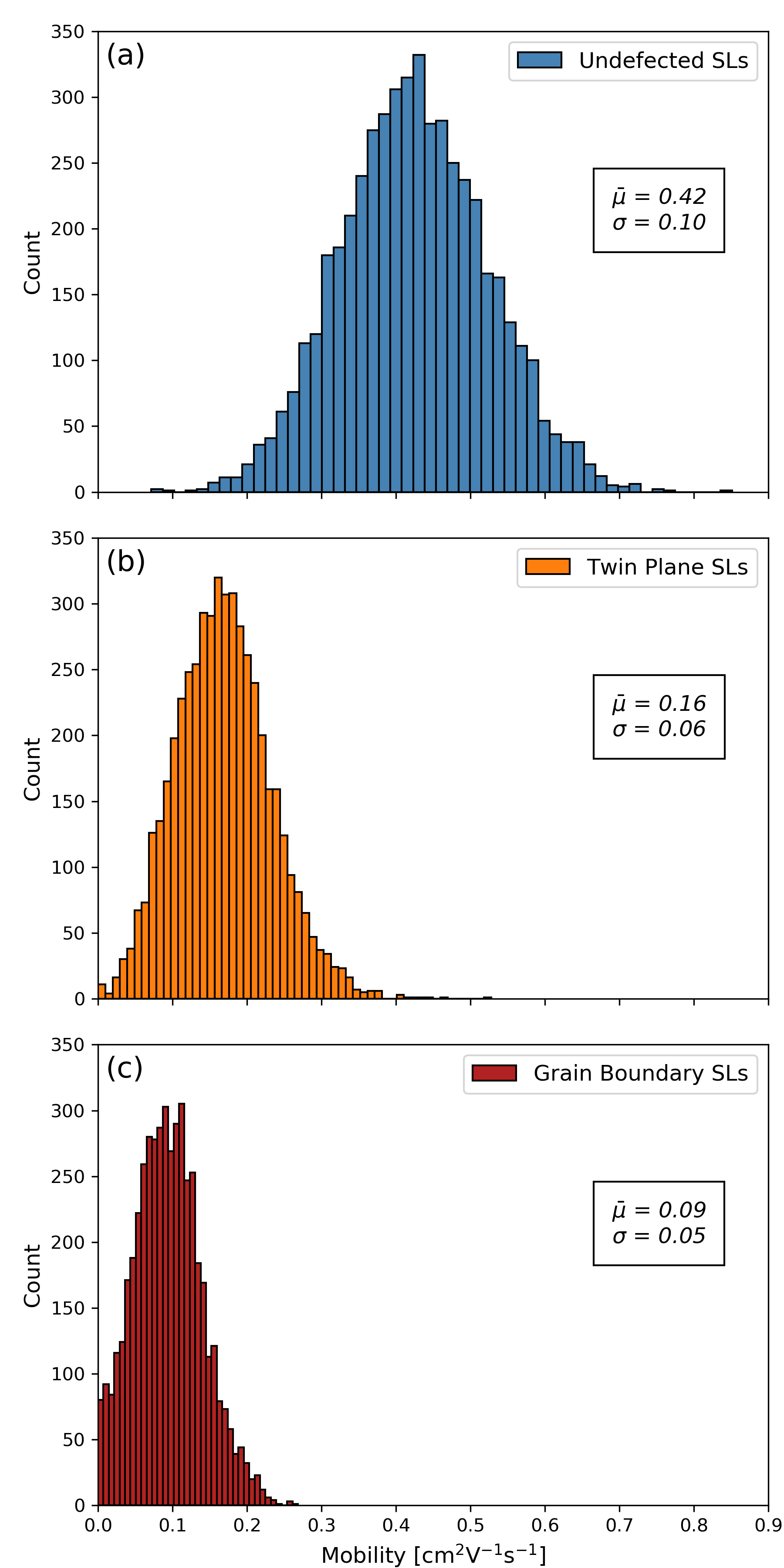}
\caption{Mobility distributions of the 3 types of NP SLs: (a) Undefected NP SLs, the NPs having size and location disorder; (b) NP SLs, each containing a twin plane, also with NP size and location disorder; (c) NP SLs, each containing a grain boundary, also with NP size and location disorder. Displayed in each panel is the average mobility and standard deviation of Gaussians fitted to the distributions. All simulations were performed at a temperature of $T=300K$.}
\label{mobilities}
\end{figure}

Next, we executed step (6) by determining the mobility distribution of the defected NP SLs. The mobility distribution for the homogeneously-disordered, undefected NP SLs is shown in Fig.\ref{mobilities}a. The mobility distribution of the twin-plane NP SLs is shown in Fig.\ref{mobilities}b. Finally, the mobility distribution of the grain-boundary NP SLs is shown in Fig.\ref{mobilities}c. All three distributions were approximately normal, and could be well characterized by a mean and a standard deviation. The mobility of the undefected NP SLs was 0.42 $\pm$ 0.1 cm$^2$/Vs, the mobility of the NP SLs containing twin planes was 0.16 $\pm$ 0.06 cm$^2$/Vs, and the mobility of the NP SLs containing grain boundaries was 0.09 $\pm$ 0.05 cm$^2$/Vs, as shown.

We then performed step (7) by assembling a resistive network whose individual links had mobilities selected from the above determined mobility distributions. To identify the paradigmatic aspects of the behavior of the mobility of the NP solid, we selected the links from the highest mobility undefected NP SL distribution with a probability $p$, and from the lowest mobility grain boundary NP SL distribution with a probability $(1-p)$. We used this $p$, the fraction of the high mobility undefected NP SLs/links as the control parameter of our study. Initially we expected that when the probability $(1-p)$ of the low mobility defected NP SLs becomes small, then the electrons will be able to ``flow around” the low mobility links through the high mobility links. Put differently, the high mobility links will be able to approximately short out the low mobility links. Had this expectation been true, then the mobility of the NP solid should have exhibited a saturation as $p$ approached 1.

\begin{figure}[h]
\includegraphics[width=0.75\columnwidth]{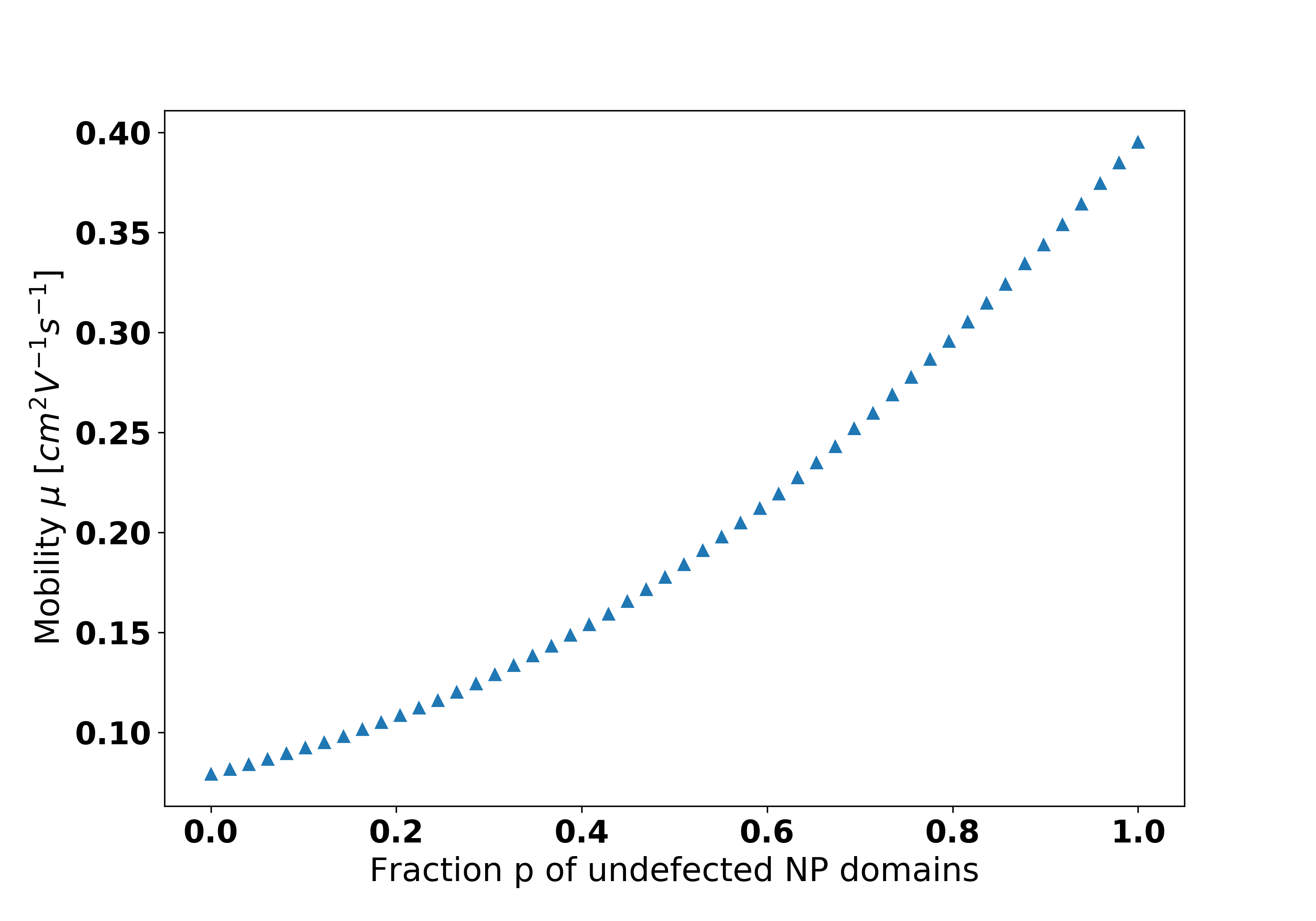}
\caption{Mobility of a 32x32 resistor network whose resistors/links are chosen from two distributions, the low mobility grain boundary NP SLs mobility distribution and the high mobility undefected NP SLs mobility distribution. The fraction p of the high mobility NP SLs sweeps the 0 to 1 region. The error bars are smaller than the data points.}
\label{mobility_with_p}
\end{figure}

Fig.\ref{mobility_with_p} shows the evolution of the mobility of NP solids with $p$. Visibly, our initial expectation was not confirmed as the mobility did not show a saturation as the fraction $p$ of the undefected NP SLs approached 1.0. The high mobility links did not ``short out" the low mobility links. A possible explanation is that the mobilities of the different NP SLs were not different enough for such a short-out. We checked the robustness of this result, and found the same characteristic non-saturating shape in other 2D geometries, as well as in 3D bulk networks. 

\underline{Bimodal percolation transition:} Next, we investigated whether there is a percolative critical behavior as the $p$ fraction of high mobility links is varied. The simple case of a resistor network, where the links either have a finite (electronic) mobility $\mu$ with probability $p$, or a non-conductive zero mobility with probability $(1-p)$, has been extensively analyzed. The conductivity of such a resistor network exhibits a critical behavior across the percolation critical point $p_c$ with a power law dependence $\mu \propto (p-p_c)^t$, where the critical exponent $t>1$ is universal, and $p_c$ is the percolation threshold.

Remarkably, the closely related bimodal problem of the links of a network having a high conductivity $\sigma(high)$ with probability $p$, or a low but finite conductivity $\sigma(low)$ with probability $(1-p)$ has been rarely analyzed. Efros and Shklovskii (ES) established the broad framework for the analysis, when they made the analogy between this bimodal distribution problem and the problem of how the critical behavior of a spin system gets modified by the presence of a symmetry breaking magnetic field\cite{shklovskii_perc}. They hypothesized a power law critical behavior for the network conductivity, where the universal scaling function at the critical point $p=p_c$ is anchored by the ratio of the high and low conductivities. However, they did not determine either the critical exponents, or the universal scaling function. 

\underline{Finite size scaling:} Next, we attempt to adapt the ES bimodal framework to describe our TRIDENS-simulated results. The finite size $L$ of the simulated samples makes it necessary to analyze the results by finite size scaling. Normally this is handled by the introduction of a scaling function with a single variable: the ratio of the sample size $L$ to the correlation length $\xi=\xi_0 P^{-\nu}$, where $P=\frac{|p-pc|}{pc}$ that smoothes over the non-analytic critical behavior.

However, for the present, bimodal mobility distribution problem ES argued that the ratio of conductivities plays the role similar to an external magnetic field in a critical magnetic system: $\frac{\sigma_{low}}{\sigma_{high}} = h$, and already smoothes over the critical behavior. Therefore, the model needs to be analyzed by a two variable finite size scaling form, where the ratio $L/\xi$ is a second factor that smoothes the critical behavior:

\begin{equation}
\mu(P,L,h)=P^{-\alpha}\mu(hP^{-\Delta},LP^\nu)
\end{equation}

where $\mu(x,y)$ is the universal finite size scaling function, and $\alpha, \nu$ and $\Delta$ are critical exponents. The analysis is more tractable if the singular $P$ dependence is absorbed by factoring out $hP^{-\frac{\alpha}{\Delta}}$ from $\mu$, leaving us with:

\begin{equation}
\mu(P,L,h)=h^{-\alpha/\Delta}\mu'(hP^{-\Delta},LP^\nu)
\end{equation}

Since $\mu'(x,y)$ is a two-variable function, the full testing of the finite size scaling hypothesis would require a quite extensive computational effort. Therefore, we narrowed our analysis of the finite size scaling assumption to the first variable, $hP^{-\Delta}$, while keeping the second variable, $LP^\nu$, constant. As the lattice sizes were varied from $L=32$ to $L=128$, keeping $LP^\nu$ constant required the appropriate modification of $P$. We then chose the $h$ values so that the critical regime on either side of the critical point $p_c$ was well sampled.

\begin{figure}[h]
\includegraphics[width=.75\columnwidth]{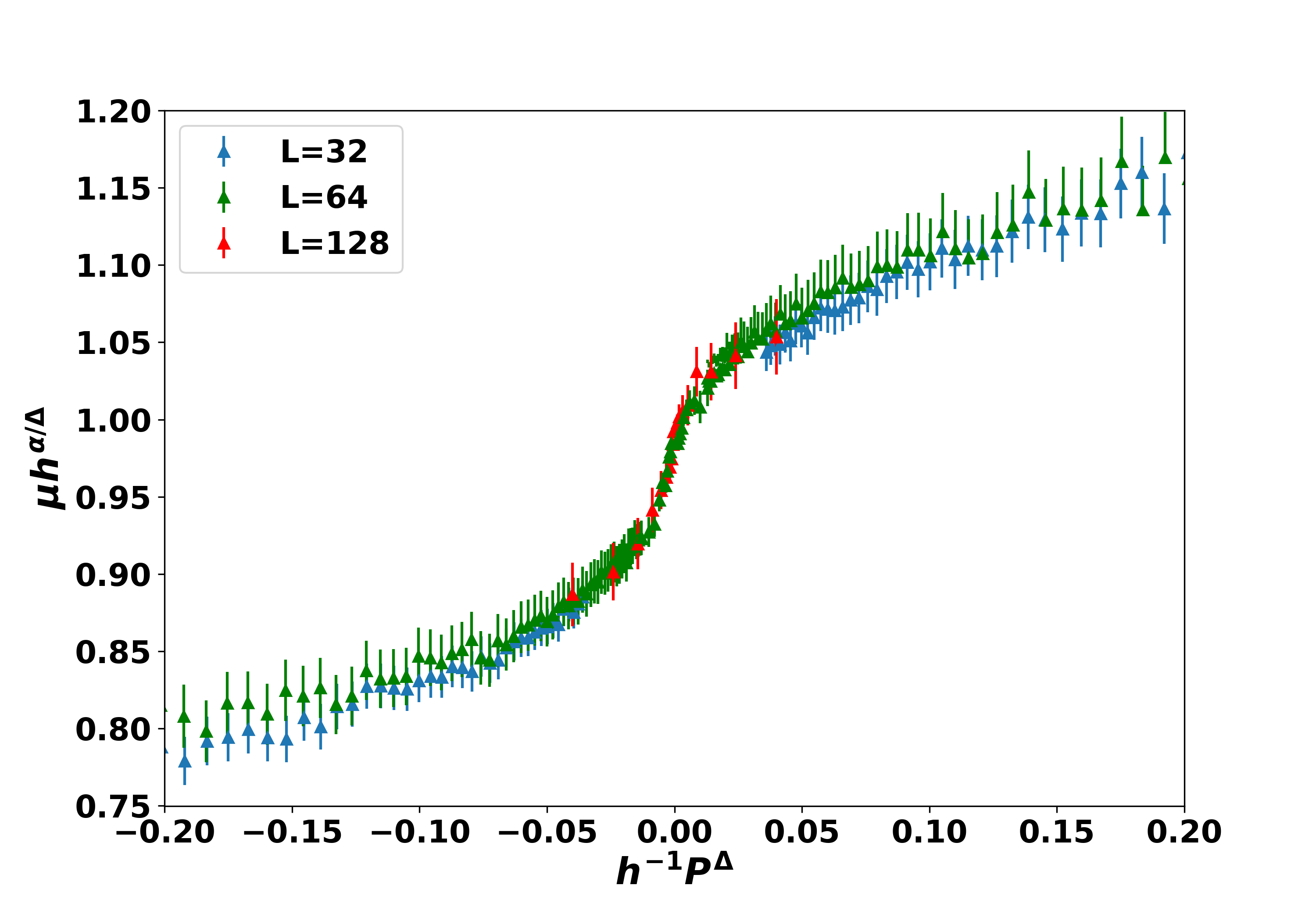}
\caption{Scaled data of $\mu h^{\alpha/\Delta}$ vs. $h^{-1}P^{\Delta}$ for 3 different lattice sizes. The product $LP^\nu$ is held constant.}
\label{scaledH}
\end{figure}

Fig. \ref{scaledH} shows the scaled mobility $h^{\alpha/\Delta}$ as a function of $h^{-1}P^{\Delta}$ for a fixed value of $LP^\nu$, for three lattice sizes, varying from $L=32$ to $L=128$. Reassuringly, we were able to achieve very good data collapse within the $(-0.1, 0.1)$ critical regime around the critical point at $P=0$, which remained acceptable out to $(-0.2, 0.2)$. Using the literature values of $\nu=\frac{4}{3}$ and $p_c=0.5$, the best collapse was reached with exponents $\alpha=-0.99\pm .02$ and $\Delta=2.02\pm .02$.

The success of the finite size scaling shows that the Efros Shklovskii analogy to critical spin systems in an external magnetic field is indeed appropriate for this bimodal percolation problem: as the fraction $p$ of the high mobility undefected NP SLs increases, one can think of the evolution of the overall mobility as a modified percolation transition, rounded by the finiteness of the mobility of the low mobility NP SLs. As far as the authors know, this is the first report of the critical exponents and the universal scaling function of the bimodal distribution resistor network problem.

The following points are worth making. Fig. \ref{scaledH} shows that the overall network electron mobility, or conductivity, displays a marked transition from a low mobility insulator behavior when the high mobility NP SLs do not percolate yet, to a high mobility insulator behavior once the high mobility NP SLs percolated. Of course, both of these regimes are insulators, so while the geometry of the NP solid undergoes a genuine percolation transition, the conductive properties exhibit only a low mobility insulator-to-high mobility insulator transport crossover, not a genuine phase transition.

We have studied a version of this problem recently on the level of our HINTS code \cite{qu2017metal}, where the NPs were connected with either low mobility activated insulating links, or high mobility, non-activated metallic links. In that version of the problem, the underlying percolation transition of the metallic links of the NP solid drove a genuine metal-insulator-transition (MIT), as the percolation of the metallic links created a genuine metallic phase. We conceptualized that MIT as a Quantum Percolation Transition, and adapted the ES bimodal mobility distribution percolation model for its description, as at any fixed temperature that version of the problem also consisted of a bimodal mobility distribution with low mobility links and high mobility links. Whether the high mobility phase is a metal or a high mobility insulator can be identified from the temperature dependence of its conductivity. In the absence of the present detailed finite-size scaling study, in Ref.\cite{qu2017metal} we developed a simple, mean-field model form for the scaling function that described the mobility's evolution from the low mobility insulator to the high mobility metal. With the notation of the present paper, the mean field exponents were $\alpha=-1$ and $\Delta=1$. This enabled us to create the dynamical phase diagram of the model on the electron filling -- disorder plane, where the MIT separated the insulating phase with activated conductivity from the metallic phase with non-activated conductivity. 

The present bimodal TRIDENS study scales up our previous bimodal HINTS work to much larger length scales. The key distinction is that the building blocks of the HINTS network were the individual NPs, whereas in TRIDENS the building blocks are the NP SLs with around a thousand NPs (in the present work, with ~800 NPs). Further, in HINTS the origin of the bimodal mobility distribution was the presence or absence of metallic links between individual NPs, whereas here the origin of the bimodal mobility distribution is the presence or absence of an planar defect across the individual NP SLs. Obviously, the HINTS transport modeling that tracks individual electrons transitiong between individual NPs is more detailed than the resistor network of the present TRIDENS work. Nevertheless, since the building blocks of both the bimodal HINTS and the bimodal TRIDENS are low mobility links and high mobility links, we expected that the same ES bimodal percolation model with the same universal scaling function and exponents will capture the critical behavior of the bimodal TRIDENS results. The success of the data collapse with the ES finite size scaling form validated this expectation. While, of course several different analytic forms can be fitted to the universal scaling function that emerged in Fig. \ref{scaledH}, nevertheless it was reassuring that in particular the mean-field function of the bimodal HINTS study:

\begin{equation}
\mu'(hP^{-\Delta},LP^\nu\rightarrow\infty)=1/(1+P/h)
\end{equation}

was also consistent with it. Further, the $\alpha=-0.99$ exponent of the TRIDENS scaling is approximately equal to the $\alpha=-1$ mean field value within the margin of error. We noted that there was a difference regarding the $\Delta$ exponent: TRIDENS gave a $\Delta=2.02$, whereas in the mean field theory $\Delta=1$. However, such differences occur typically between mean-field and numerically determined exponents. All in all, the substantial correspondence between our HINTS and TRIDENS works demonstrated that the ES bimodal percolation model is a good, quantitative description of how the underlying percolation transition of the NP solid drives a low-mobility insulator-to-high-mobility insulator transport crossover.

For completeness we mention that we implemented TRIDENS with randomly selecting high or low mobility NP SLs for the links. This corresponds to planar defects with a length of tens of NPs. However, the sample in Fig.\ref{sem} has many defects that are much longer. Such long defects can be modelled by selecting defected SLs along lines of links in TRIDENS, in a correlated manner. Such correlated TRIDENS models will be pursued in a future work.

\begin{center}
\underline{Conclusions}    
\end{center}

Transport in nanoparticle solids must be simulated on extremely large length scales, corresponding to millions of NPs, because NP solids exhibit spatial structures on several length scales, from the subtleties of individual NPs through the sensitive modeling of inter-NP transitions and through transport across homogeneously disordered SLs all the way to transport in NP SLs with large planar defects. Single-level computational methods are manifestly unable to span these length scales. This is why in our previous work we developed the multi-level HINTS method that was capable of simulating transport across NP solids with up to a thousand NPs. However, even HINTS is unable to capture the effect of planar defects on transport in NP solids of the size of tens of microns.

In this paper, we reported the development of the TRIDENS method that adds three further hierarchial layers on top of the HINTS method. In TRIDENS, we first introduced planar defects into individual NP SLs that comprised the order of about a thousand NPs. Then we used HINTS to simulate the transport across these defected NP SLs. We performed these HINTS transport simulations for tens of thousands of defected NP SLs, and constructed the distribution of the NP SL mobilities with planar defects. Second, the defected NP SLs were assembled into a resistor network with more than $10^4$ NP SLs, thus representing about $10^7$ individual NPs. This translated to length scales of tens of microns, approaching the experimental scales for NP solids. Third, and finally, the TRIDENS results were analyzed by finite size scaling to explore whether the percolation transition, separating the phase where the low-mobility-defected NP SLs percolate from the phase where the high-mobility-undefected NP SLs percolate, drives a low-mobility-insulator-to-high-mobility-insulator transport crossover that can be extrapolated to genuinely macroscopic length scales.

Our extensive TRIDENS analysis generated the following results. On the level of individual NP SLs, we found that the average of the mobility for undefected NP SLs was 0.42 $\pm$ 0.1 cm$^2$/Vs, for twin-plane-defected NP SLs 0.16 $\pm$ 0.06 cm$^2$/Vs, and for grain-boundary-defected NP SLs 0.09 $\pm$ 0.05 cm$^2$/Vs. On average, grain boundary defects hinder transport about twice as much as twin planes. This result makes sense, as grain boundaries are more disruptive to lattice periodicity than twin planes, and transport across the grain boundaries involves longer hops between more distant NPs, whereas transport across twin planes proceeds across many NPs shared by the grains on the two sides of the twin plane, and thus it involves regular hop lengths. It is noteworthy that the introduction of planar defects into NP SLs reduced their mobility by a factor of up to 5. On one hand, this is a substantial suppression of the mobilities that drives a transport crossover, and thus demonstrates the imperative of minimizing the density of planar defects in NP solids to help their suitability for applications. On the other hand, this is not a qualitative, order-of-magnitude suppression of the transport that indicate a Metal-Insulator-Transition: those are driven by the loss of phase coherence.

On the highest, resistor network-level analysis of TRIDENS, we observed that the introduction of the planar defects immediately started to reduce the network mobility. This finding suggests that even small concentrations of planar defects are not shorted out in NP solids, and thus every reduction of the density of planar defects will lead to further improvements of the transport in NP solids. Among the planar defects, the elimination of grain boundaries pays more dividends than that of twin planes.

For the theoretical description, we adapted the Efros-Shklovskii bimodal mobility distribution percolation model. We performed a finite size scaling analysis of the TRIDENS network mobilities. We demonstrated that increasing the density of the undefected NP SLs drives an underlying, structural/geometric percolation transition in the NP solid, which in turn drives a low-mobility-insulator-to-high-mobility-insulator transport crossover. We demonstrated that our adaptation of the ES bimodal theory's two-variable scaling function is an effective tool to quantitatively characterize this low-mobility-insulator-to-high-mobility-insulator transport crossover. For context, we discussed the analogies with the Quantum Percolation Transition we developed in our earlier, MIT-focused work\cite{qu2017metal}.

\underline{Acknowledgements}. C.Q., A.A., and M.L. were supported by the UC Office of the President under the UC Laboratory Fees Research Program Collaborative Research and Training Award LFR-17-477148. Materials characterization was performed at the user facilities of the UC Irvine Materials Research Institute (IMRI). C.H., D.U. and G.T.Z. were supported by the National Science Foundation under award DMR-2005210.

\bibliography{kmc_bib.bib}

\end{document}